\documentstyle[14pt]{article}
\begin{document}
\textwidth 16cm
\textheight 23cm
\newcommand{\gma}{\mbox{$\gamma^{\,\prime}$}}
\begin{center}
     {\large\bf The glueball Regge trajectory from the string -
     inspired theory}\\[5mm]
     S.V. Talalov\\
     {\small Dept. of Theoretical Physics,  University
      of Togliatti, \\
    13 blvrd Korolyova, Togliatti, 445859, Russia}\\[5mm]

\begin{minipage}{13cm}
\paragraph~
{\small  The special case of $4D$ string-like theory proposed
     early is investigated. Regge trajectories in the developed
model are  non-linear for the small masses and the values of spin and
     have the different asymptotical slopes $\alpha_p$. The
value $\alpha_p$ just as the form of the trajectory depends from the
quantum state of some ''internal`` (relativistic invariant) variables.
   It is also shown that some trajectories have the
    asymptotical slope $\alpha_g\simeq 0.21 Gev^{-2}.$}
    \end{minipage}
     \end{center}

     \subsection*{I. Introduction}
     \paragraph~

     The conventional approaches in a string theory
     (see, for example, \cite{GSW,BarNes})
     lead as is well known to the linear Regge trajectories for free
     string such that the slope $\alpha^\prime$ is in-put parameter
     in a theory.  But
     the   trajectories $s=\alpha^\prime M^2+c,$
     where the value $\alpha^\prime\simeq 0.9 Gev^{-2}$ is the universal
     constant, describe the spectrum of  real particles well
     but only approximately. Indeed, the linearity means that the width
     of any resonance is equal to zero;
     the universality of the slope $\alpha^\prime$ is
     connected with the absence of  exotic particles
     \cite{DeAlf}.
     In the meantime we have the stable evidence on hadronic
     exotics now
     \cite{Landsb}.
     These data  give the true information about Regge trajectories
     with slopes $\alpha_g\not= 0.9 Gev^{-2}$
     \cite{Igi,HoSo,FiSr}.
     As regards  the form of the trajectory, the linear
     dependence gives a good approximation for light-flavoured mesons
     and baryons only (see, for example
     \cite{Serg}). Therefore  we have to modify standard string theory
      to describe the real (extended) particles.
     Of couse, such modification must solve in some way the problem of
     the extra space - time dimensions  which is usual for conventional
     approaches. To solve this problem the various   (super)string models
     have been constructed by  many authors both in the physical ($4D$)
     and in the arbitrary (non-critical) dimensions (see, for example,
  \cite{Roh, Poly, GerNev, Mar2, Chams, Pron, Lun}.)\footnote{This list
     can be  continued, of couse.}.
     Most of them lead to the linear Regge trajectories too.

     Resently new approach was suggested to describe the classical and
    quantum dynamics of $4D$ open spinning string
     \cite{Tal1}.
     Proposed theory differs, in our knowledges, from other
     because it  founded on the new conception of ''adjunct phase space``.
     Generally speaking, we describe the classical dynamics of the
     string with canonical phase space ${\cal H}$,  fist type
     constraints $F_i$ and some additional conditions $G_i$
     in terms of another dynamical system
     with phase space ${\cal H}^{ad}$. These phase spaces
     connected in accordance with diagramm
     \begin{equation}
     \label{diagra}
     {\cal H}\supset{\bf V}\approx
     {\bf W}\subset{\cal H}^{ad},
     \end{equation}
     where set ${\bf V}$ is the surface of the constraints and
     additional conditions in the  space ${\cal H}$,
     set ${\bf W}$ is the surface of some (first type) constraints
     $\Phi_i$, $(i=1,\dots,n<\infty)$
     in the  space ${\cal H}^{ad}$. The symbol  $\approx$
     means the diffeomorfism  conserved in the dynamics.
     From the classical viewpoint, the manifold ${\bf W}$
     is equivalent to the manifold ${\bf V},$ because it
     contains same information about the physical degrees of the
     freedom. It should be stressed that ${\cal H}\not\approx{\cal H}^{ad}$
     as the Poisson manifolds: there is no any canonical transformation
     ${\cal H}\to{\cal H}^{ad}$.
     We  fulfill the subsequent quantization of the string theory
     in terms of the space ${\cal H}^{ad}.$
     Rigorously speaking, we quantize  another dynamical system and
     select the ''string sector`` which corresponds to the set
     ${\bf W}.$  Relativistic invariance does not broken in
     quantum case because special selection of the phase space
     ${\cal H}^{ad}$ which will be quite natural in our opinion.
     Finally, as a quantum theory we have the operator representation
     of the classical  algebra
     $${\cal A}_{cl}={\cal P}\oplus{\cal A}_{int},$$
     where ${\cal P}$ is the Poincar\'e algebra and
     ${\cal A}_{int}$ the Poisson brackets algebra of some
     two-dimensional fields in the ''box`` (there are
     ''internal`` variables).
     Thus our approach is a natural  generalization of the Wigner's
     conception of ''elementary`` particle as the representation of
     Poincar\'e algebra. Non-trivial spectrum appears here because
     existence of constraints depending from the Cazimir functions
     $P^2$ and $w^2$.
     We interpret such quantum theory as the model of extended
     relativistic particle.

     \subsection*{II. Basic points of a classical theory.}
     \paragraph~

    As the model of spinning string  we consider the following
    theory. Let
     the fields $X_\mu(\xi^0,\xi^1)$ and
     $\Psi^A_{\pm}(\xi^0,\xi^1)$
     interact with two-dimensional gravity
     $h_{ij}(\xi^0,\xi^1),$
    where $\xi^1\in[0,\pi]$ and $\xi^0\in(-\infty,\infty),$
    such that dynamics is defined by the action
      constructed in accordance with the
     well-investigated manner
     \cite{GSW}:
     \begin{equation}
     \label{action}
     S=-{1\over{4\pi\alpha^{\prime}}}\int d\xi^0d\xi^1\sqrt{- h}
     \left\{h^{ij}\partial_i X^\mu\partial_j X_\mu -
     i{\varrho_0} e^j_I(\Gamma^0)_{AB}\overline\Psi^A\gamma_j
     {\nabla^I}\Psi^B\right\}.
     \end{equation}
     The notations are  following:    $h=\det(h^{ij})$,
     the vectors  $e^j_I(\xi^0,\xi^1)$  are the vectors of two-dimensional
     basis such that the equalities  $h^{ij} = e^{iI}e^j_I$ take place
     and the matrices  $\Gamma^\mu$  and $\gamma_i$ are the Dirac
     matrices in the four- and two-dimensional space-time respectively.
     The field  $X_\mu$ takes the values   in
     Minkowski space-time  $E_{1,3};$  the fields  $\Psi^A$
     with components $\Psi^A_\pm$ are the spinor fields in two-dimensional
     space; index $A$ is the spinor index in the space $E_{1,3}$
      such that the fields  $\Psi_\pm$ are the Majorana spinor
     fields in four-dimensional space-time.
     The numbers $\Psi^A_\pm$ are the complex numbers, thus
     there are no classical Grassmann variables in our action.
     The consideration of the spinning string without the
     Grassmann variables is not new (see, for example,
     \cite {Bar}).

     To fix the gauge arbitrariness  we demand, as usually,
     $e^j_I=\delta^j_I$ so that $h_{ij}={\rm diag}(1,-1)$ and
     the equations of motion can be written in simplest form.
     For fields $X$ and $\Psi$ we have
     $\partial_-\partial_+X_{\mu}=0,$
     $\quad\partial_{\mp}\Psi_{\pm}=0;$
     the equations of motion $\delta S/\delta h^{ij}=0$
     for gravity $h$ lead as well-known to the equalities
     \begin{equation}
      \label{prcon1}
     F_{\pm}(\xi)\equiv
     \left(\partial_\pm X\right)^2\pm{{i\varrho_0}\over 2}\,
     \overline\Psi_\pm\partial_\pm\Psi_\pm=0,
     \end{equation}
     where $\partial_\pm$ are derivatives with respect to
     cone parameters $\xi_\pm=\xi^1\pm\xi^0.$

     We still have the remained gauge freedom
     \cite{GSW}
     \begin{equation}
     \label{coninv}
     \xi_\pm\longrightarrow\tilde\xi_\pm=\pm A(\pm\xi_\pm).
     \end{equation}
     The  function $A(\xi)$ must satisfy  the property
     $A(\xi+2\pi)=A(\xi)+2\pi$~ $(A^\prime\not=0)$
     in accordance with the standard boundary conditions for
     original variables $X$ and $\Psi$:
     $ X^{\prime}_{\mu}(\xi^0,0)= X^{\prime}_{\mu}(\xi^0,\pi)=0,$~
     $\Psi_+(\xi^0,0)=\Psi_-(\xi^0,0)$ and
     $\Psi_+(\xi^0,\pi)= \epsilon\Psi_-(\xi^0,\pi),$
      where $\epsilon=\pm.$
     Our subsequent studies are founded on two conjectures.
     Fistly, we restrict the set of all string configurations
     by requirement
     \begin{equation}
     \label{restric}
     \pm\overline{\Psi}_{\pm}\Gamma^{\mu}\Psi_{\pm}
     \partial_{\pm}X_{\mu}>0\,.
     \end{equation}
     Secondly, we introduce the additional conditions to fix
     all degrees of the gauge freedom (\ref{coninv})
     excepting  one. We suppose  that the equalities
     \begin{equation}
     \label{gauge}
     \partial_{\pm}\left[\overline{\Psi}_{\pm}\Gamma^{\mu}
     \Psi_{\pm}\partial_{\pm}X_{\mu}\right]=0\,.
     \end{equation}
     hold.
     It is more convenient to integrate eq. (\ref{gauge})
     so that
\begin{equation}
     \label{addcon}
     G_{\pm}(\xi)\equiv
\overline{\Psi}_{\pm}\Gamma^{\mu}\Psi_{\pm}\partial_{\pm}X_{\mu}
=\pm{\kappa^2\over 2}
\end{equation}
for any positive constant $\kappa.$ These equalities are equivalent
to the original conditions (\ref{restric}) and (\ref{gauge}).
     Note that the conditions (\ref{addcon}) are
     invariant both under Poincar\'e and under scale transformations
     of the space-time $E_{1,3}.$
      Such invariance is first reason of the motivation for the conditions
    (\ref{addcon}).  Second reason is to the  gauge  (\ref{addcon})
     generalizes naturally the well-known light-cone gauge in a
     string theory ( this fact is discussed firstly in the work
     \cite{Tal2} in connection with geometrical description $3D$
     spinning string).
     We define the set ${\bf V}$ as a set of the string configurations
     which satisfy the constraints (\ref{prcon1}) and the additional
     conditions  (\ref{restric}) and (\ref{gauge}).

      The original phase space ${\cal H}$ has the coordinates
     $\dot{X}_\mu\equiv\partial_0 X_\mu$, $X_\mu$,
      ${\mathop{\Psi^+}^A}_\pm$ and $\Psi^A_\pm.$ As usually,
     canonical Poisson bracket structure is following:
     $$\{\dot{X}_\mu (\xi),X_\nu (\eta)\}=-4\pi\alpha^\prime
     g_{\mu\nu}\delta(\xi-\eta),$$
     $$\{{\mathop{\Psi^A}^+}_\pm(\xi),\Psi^B_\pm(\eta)\}=
     {8\pi i\alpha^\prime\over{\varrho_0}}(\Gamma^0)^{AB}\delta(\xi-\eta).$$

     Let us define two functions $F$ and $G$ which will be continious
     and $2\pi$-periodical in accordance with boundary conditions
     on original string variables:

     $$F(\xi)=\cases{F_{+}(\xi),\cr
     F_{-}(-\xi),\cr}\quad,\quad
     G(\xi)=\cases{G_{+}(\xi),& $\xi\in[0,\pi)$,\cr
     G_{-}(-\xi),& $\xi\in[-\pi,0)$,\cr}\,.$$

     In terms of  Fourier modes $F_n$ and $G_n$ of the introdused
     functions $F$ and $G$, the set ${\bf V}$ is defined by
     the equalities
     $$ F_n=0 \quad (n=0,\pm 1,\dots),\qquad
     G_n=0 \quad (n\not=0)\,.$$
     In accordance with canonical Poisson bracket structure
     we have the equalities
     $$\{{\mathop{F}^*}_n,G_m\}=4\pi i\alpha^\prime
     (n+m)G_{n-m}\,. $$
     Because  $G_0=\kappa^2/2>0$ in our theory,
     the  system of ''constraints``
     \begin{equation}
     \label{FGcons}
      F_n=0,\quad (n\not=0),\qquad
     G_n=0 \quad (n\not=0)
     \end{equation}
    will be the  second type system in Dirac terminology.
    We can  introduce correspondent Dirac brackets and
    consider the reduced phase space ${\cal H}_1$
    (defined by the equalities (\ref{FGcons})).
     In this space the set ${\bf V}\subset{\cal H}_1$
     will be the surface of single constraint
     $F_0=0.$ It is clear that such constraint
     generates the transformations (\ref{coninv}) such that
    $A(\xi)\equiv\xi+c,$ where $c=const.$ Obviously,
     they give the shifts $\xi^0\rightarrow\xi^0+c,$ which
     correspond to $\xi^0$-dynamics.

     As a next step of our classical theory we consider
     another dynamical hamiltonian system (''extended particle``).
     To do it we must define the objects:
     a phase space  ${\cal H}^{ad}$, a Poisson brackets
      $\{\cdot,\cdot\}^0$,  a  hamiltonian $h_0$,
     and, may be, some constraints $\Phi_i$.
     The definitions will be following.
     The space  ${\cal H}^{ad}$ has a structure
     $${\cal H}^{ad}=
     {\cal H}_{p}\times{\cal H}_b\times{\cal H}_j\times{\cal H}_0,$$
     where:
     ${\cal H}_{p}$ is the algebraic space of the Poincar\'e
     algebra with coordinates $P_\mu$ and  $M_{\mu\nu}$;
     ${\cal H}_{b}$ -- the phase space of the complex D'Alembert field
     $b(\xi^0,\xi^1)$ which is defined for  $\xi^1\in[0,\pi]$ and
     satisfied to the boundary conditions
      $$b(\xi^0,0)=b(\xi^0,\pi)=0;$$
     the space  ${\cal H}_{j} = {\cal H}_{j_0}\,\times\,{\cal H}_{U}$
     be composed from the space
     ${\cal H}_{j_0}$ which is the phase space of the real D'Alembert
     field  $j(\xi^0,\xi^1)$ in same ''box``   $\xi^1\in[0,\pi]$
     and satisfied to the boundary conditions
     $$j(\xi^0,0)= 0,\qquad j(\xi^0,\pi)= 2\pi j_0^0$$
     and from the space  ${\cal H}_U$. Last one is the phase space of
     $SU(2)$ - valued WZNW - field $U(\xi^0,\xi^1)$ defined in the
     ''box`` $\xi^1\in[0,\pi]$ and satisfied to the boundary conditions
     $$U(\xi^0,0)= U(\xi^0,\pi)= 1_2.$$
     The space ${\cal H}_0$ is some auxiliary (finite - dimensional)
     space; it is not important here and will be dropped out from
    subsequent consideration.
     We use following representations for introduced fields:
         $$b(\xi^0,\xi^1)=\int^{\xi_+}_{-\xi_-}f(\eta)d\eta,$$
     where $f(\xi)$ -- complex $2\pi$ - periodical function
     without zero mode;
         $$j(\xi^0,\xi^1)=\int^{\xi_+}_{-\xi_-}j_0(\eta)d\eta,$$
     where  $j_0(\xi)$ -- real $2\pi$ - periodical function
     such that the zero mode of this function is the constant $j_0^0$.
     The space  ${\cal H}_U$  is coordinatized by three real functions
     $j_a(\xi)$ $(a=1,2,3)$ (see  \cite{TakFad}).
     $2\pi$ - periodicity of these functions is the consequence of the
     boudary conditions  for the WZNW-field $U$.
     We assume that all coordinates of spaces
     ${\cal H}_{b}$,  ${\cal H}_{j}$ and  ${\cal H}_{0}$
     are Poincar\'e - scalars.

     Natural Poisson brackets for the coordinates of the phase space
      ${\cal H}^{ad}$ will be following\footnote{Because the functions
     $f$ and $j_a$ are dimensionles we must use  some  constant  $S_0$
    which has a dimension of action. For the subsequent calculations
    $S_0=\varrho^2_0/\alpha^{'}$ and $c=\hbar=1.$}:
    \begin{eqnarray}
   &\left\{M_{\alpha\beta},M_{\gamma\delta}\right\}^{0} &=
g_{\alpha\delta}M_{\beta\gamma}+g_{\beta\gamma}M_{\alpha
\delta}-g_{\alpha\gamma}M_{\beta\delta}-
g_{\beta\delta}M_{\alpha\gamma},\nonumber\\[3mm]
   &\left\{M_{\alpha\beta},P_\gamma\right\}^{0} &=
g_{\beta\gamma}P_{\alpha}-g_{\alpha\gamma}P_{\beta},\qquad
   \left\{P_{\alpha},P_\gamma\right\}^{0}= 0,\nonumber\\[3mm]
    & \left\{f(\xi),\overline f(\eta)\right\}^{0} &=
      {S_0^{-1}}\delta^{'}(\xi-\eta),\nonumber\\[3mm]
    &\left\{j_0(\xi),j_0(\eta)\right\}^{0} &=
     -2{S_0^{-1}}\delta^{'}(\xi-\eta),\nonumber\\[3mm]
 \label{jajb}
     &\left\{j_a(\xi),j_b(\eta)\right\}^{0} &=
     2{S_0^{-1}}\left(-\delta_{ab}\delta^{'}(\xi-\eta) +
      \varepsilon_{abc}j_c(\xi)\delta(\xi-\eta)\right).
     \end{eqnarray}
    ($a,b,c=1,2,3$ and    $\delta(\xi)=\sum_n{\rm
    e}^{in\xi}$ here.)  As hamiltonian we postulate the function
     $$h_0={S_0\over{2\pi}}\left(\int_0^{2\pi}\vert f(\xi)\vert^2
     d\xi+{1\over 4}\sum_{a=0}^3
     \int_0^{2\pi}j_a^2(\xi)d\xi\right),$$
     so that the following non-trivial equations of motion hold:
     \begin{equation}
     \label{3.eqmo1}
     \left\{h_0,j_a\right\}^0 = j_a^\prime,\qquad
     \left\{h_0,f\right\}^0 = f^\prime\,.
     \end{equation}

     {\bf Main result} of the work
     \cite{Tal1}
     is formulated as the following statement.

     \begin{em}
     The equality
     \begin{equation}
     \label{mainco}
     \Phi(P^2,w^2;f,j_0,\dots,j_3)=0,
     \end{equation}
     where $w^2$ is the square of the pseudo-vector
     $w_{\mu}=-(1/{2})
     \varepsilon_{\mu\nu\lambda\sigma}M^{\nu\lambda}P^{\sigma}$,
      exists on the phase space ${\cal H}^{ad}$
     such the surface ${\bf W}\subset{\cal H}^{ad}$, corresponded
     to ''constraint`` (\ref{mainco}) will be diffeomorfical to the
     surface ${\bf V}$ in the string phase space ${\cal H}$.
     Coordinates $P_\mu$ and $M_{\mu\nu}$ of the space
     ${\cal H}^{ad}$ corresponded any string configuration
     $(X_\mu,\Psi_\pm)$ for such diffeomorfism, will coincide
     with the string energy-momentum and moment calculated
     for this configuration in accordance with the Noether theorem.
     Constraint (\ref{mainco}) is stable, i.e. the equality
       $$\{h_0,\Phi\}^{0}=0$$
     takes place.
     \end{em}

     Formulated statement means that  the inclusions
     \begin{equation}
     \label{diagr1}
     {\cal H}_1\supset{\bf V}\approx{\bf W}\subset{\cal H}^{ad}
     \end{equation}
     are fulfilled.
     We use  the redused phase space ${\cal H}_1$
     instead the  phase space ${\cal H}$ here to demonstrate
     the connection of the string  with the dynamical system ''extended
     particle``.  In our opinion, it  clarify suggested theory.
     Indeed, ${\rm codim}{\bf V}=1$ in the space ${\cal H}_1$
     as ${\rm codim}{\bf W}=1$ in the space ${\cal H}^{ad}$.
     Thus we can interpret our approach as such  deformation of the
     phase space ${\cal H}_1$  that the physical degrees of the freedom
     (set ${\bf V}$) are non-deformed.

     The  ''constraint``  (\ref{mainco}) was deduced in the
     work \cite{Tal1} in  the $\kappa$-parametric form:
     \begin{eqnarray}
     \label{P2}
     P^2&=&\left({\varrho_0\over{4\pi\alpha^\prime}}\right)^2
     \sum_{l=0}^2\left(\kappa\over\varrho_0\right)^l [C_p]_l,\\
     \label{w2}
     w^2&=&{\varrho_0^6\over({4\pi\alpha^\prime})^4}
     \sum_{l=0}^6 \left(\kappa\over\varrho_0\right)^l [C_w]_l.
     \end{eqnarray}
     The evident expressions for the coefficients $[C_p]_l$ and
     $[C_w]_l$ in the polynomials (\ref{P2}) and (\ref{w2})
     are defined  by the
     Noether formulae for string energy-momentum and moment.
     It is important that the coefficients $[C_p]_l$ and $[C_w]_l$
      depend on the functions $f(\xi)$ and $j_a(\xi)$ only.
     The equalities (\ref{P2}) and (\ref{w2}) demonstrate the
     general form of the classical Regge trajectories in our theory.
     It is clear that the asymptotical dependence  $\sqrt{S^2}\propto P^2$
     takes place for the large values $P^2$ and $S^2=w^2/P^2$.
     The exclusion of the parameter $\kappa$ from the equalities
     (\ref{P2}) and (\ref{w2})  gives the closed form (\ref{mainco}).
     The evident expression of the function $\Phi(\dots)$ as its
     domain must be described more detail in general. We will
     consider reduced case here and prefer the form
     (\ref{P2}) and (\ref{w2}). Let us note that the continuation of
     these equalities on the domain $\kappa\le 0$ is quite possible.
     Although such domain does not correspond any string, it may be
     interesting too.

     \subsection*{III. Reduction and quantization}
     \paragraph~

      The  variables of the phase space ${\cal H}^{ad}$
     are complicated functions from the string fields $X$ and
     $\Psi,$  that is why the correct selection  of quantum statistics
     for any ''internal`` variables $f$ and $j_a$ is not so obvious here.
     The following proposition clarifies
     this question \cite {Tal3}

     \begin{em}
      { The equalities $\Psi^A_\pm(\xi)\equiv{\rm const}$ hold
     if and only if the equalities $j_a(\xi)\equiv 0$ for
      $a=0,\dots,3$ take place.}
     \end{em}

     This statement means that, in spite of the complicated dependence
     of the variables $f$ and $j_a$ from the  variables
     $X_\mu$ and $\Psi,$ the bosonic and fermionic degrees of the
     freedom are still non-mixed. It is natural to fulfill the
     quantization of the variable $f$ in terms of some bosonic field
     but the variables $j_a$ in terms of the fermionic fields
     with help of the bosonization procedure
     \cite {Wit}. Thus the natural Hilbert space of the
     quantum states in our theory will be the following:
     $${\bf H}=\mathop\bigoplus_{l,i,s}\Bigl({\bf H}_b\otimes
     {\bf H}_j\otimes{\bf H}_{\mu^2_i,s}\Bigr),$$
     where the spaces ${\bf H}_{\mu^2,s}$ are the spaces
     of irreducible representations of Poincar\'e algebra
     ${\cal P},$ corresponded to the eigenvalues $\mu^2$ and
     $s(s+1)$ of the Cazimir operators  $\alpha^{'}P^2$ and $S^2;$
     ${\bf H}_b$ -- the Fock space of two-dimensional complex  bosonic
     field in the ''box`` and ${\bf H}_j$ -- the Fock space of
     some two-dimensional fermionic field in the ''box``.
     The corresponding physical  states $\mid\psi_{phys}\rangle$ must
     being selected  with help of the ''Shr\"odinger equation``
     \begin{equation}
     \label{Shrod}
     \Phi(P^2,w^2,\dots)\mid\psi_{phys}\rangle=0,
     \end{equation}
     where $\Phi_i$ are the quantum expressions for
     constraint (\ref{mainco}).

     Thus the reduction $j_a(\xi)\equiv 0$ corresponds to
     open  $4D$ string with additional spinor field which has a
     constant non-zero components  on the world-sheet.
     {\it We consider in detail the quantum version of the dynamical
     system ''extended particle`` for such reduction in this work.}
     Let us introduce the following objects.
     \begin{itemize}
     \item  bosonic creation and annihilation operators
     ${\bf b}^+_n$, ${\bf b}_n$  $(n=\pm1,\pm2,\dots)$ acting in
     the corresponding Fock space ${\bf H}_b$ such that the
     canonical commutation relations
      $[{\bf b}_n,{\bf b}^+_n]=\delta_{mn}$ hold;
     \item
     some sequence of non-negative numbers
     $$\{\mu^2_i(s)\} \qquad (s=0,1/2,1,3/2,\dots,\qquad
     i={0,1,\dots,\infty})\,.$$
     \end{itemize}
     All quantum states in our theory are the vectors
     of space
     $${\bf H}=\mathop\bigoplus_{i,s}\,\Bigl({\bf H}_b\otimes
     {\bf H}_{\mu^2_i(s),s}\Bigr)\,.$$
     It should be stressed that we can drop out the summands which
     correspond to integer or half-integer values of the number $s$.
     Therefore we can construct the models both bosons and fermions
     which have the additional bosonic internal degrees of the
     freedom.
     It is interesting to note that we can include in the definition of the
     space ${\bf H}$ the summands ${\bf H}_b\otimes {\bf H}_{i\mu}$
     which  correspond to the irreducible representations of the
     Poincar\'e  algebra with $\mu^2<0$. This case corresponds to the
     theory with the tachyon states. In accordance with our
     definition $\mu^2\ge 0$ so  there are no tachyons in the model.

     {\it We find the sequence   ${\mu^2_i(s)}$ such that
      non-zero solutions $|\psi_{phys}\rangle\in{\bf H}$
     of the ''Sr\"odinger equaton``   (\ref{Shrod}) exist and
     have the finite norm.}

     Let us define the operator-valued function
     $${\bf f}(\xi)=\sum_{n>0}\sqrt{n}\left({\bf b}_{-n}{\rm e}^{-in\xi}+
     {\bf b}^+_n{\rm e}^{in\xi}\right).$$
        It is clear that the equality
     $$\left[{\bf f}(\xi),{\bf f}^+(\eta)\right]=
      i\delta^{'}(\xi-\eta)$$
     takes place.

     To construct the quantum theory correctly, we must describe
     evidently the algebra of the classical dynamical variables
     (''observables``). By definition the observables will be  the
     functions:
     $${\cal D}={\cal D}[f(\xi);P_\mu,M_{\mu\nu}] \equiv
     {\cal D}^m[A_1,A_2,\dots,A_n]\,,$$
     where ${\cal D}^m$  some m-th power polynomial of the
     variables $A_i$. Last ones  can be the following quantities:
     \begin{enumerate}
     \item  the component of the energy-momentum $P_\mu$;
     \item  the component of the moment $M_{\mu\nu}$;
     \item  some integral
      $$  \int_0^{2\pi}\dots\int_0^{2\pi}\varphi(\xi_1,\dots,\xi_k)
      f(\xi_1)\dots f^{'}(\xi_i)\dots f^+(\xi_k)d\xi\,,$$
      such that the kernel $\varphi\in C^{\infty}$ non-degenerated.
      \end{enumerate}
      The polynomial coefficients are arbitrary analytical
     functions from the values  $P^2$ and $w^2$.

      The quantization  $\Upsilon$ for considered reduced
     dynamical system will be the correspondence
      $${\cal D}\rightarrow\Upsilon({\cal D})
      \equiv {\cal D}[\sqrt{\gma}{\bf f}(\xi); {\bf P}_\mu,
     {\bf M}_{\mu\nu}],$$
     where the number $\gma=\alpha^{'}\hbar/\varrho_0^2$ is the natural
     dimensionless parameter in our theory and bold-faced symbols
     ${\bf P}_\mu$ and ${\bf M}_{\mu\nu}$ mean the correspondent
     operators acting in the spaces ${\bf H}_{\mu^2,s}$.
     Ordering rules are following:
     \begin{enumerate}
     \item  in the integrals
     $$\Upsilon\biggl(f(\xi)\dots f^{'}(\xi)\dots f^+(\xi)\biggr)=
      :{\bf f}(\xi)\dots{\bf f}^{'}(\xi)\dots{\bf f}^+(\xi): \quad ;$$
     \item in the polynomial ${\cal D}^m$
     $$\quad\Upsilon(A_1\dots A_k)={1\over k!}\sum_{i_1,\dots,i_k}
      \Upsilon(A_{i_1})\dots\Upsilon(A_{i_k})\,,$$
     where the summands correspond to the  rearrangements of the
     numbers $1,\dots,k.$
     \end{enumerate}
     Thus all observables in our model are well-defined
     operators in the space ${\bf H}$
     (the operator ${\bf b}_n$ has to be defined as
      $\oplus_{i,s}({\bf b}_n\otimes1)$ and so on).
     The constraint  $\Phi=0$ leads to the equation (\ref{Shrod});
      solutions of this equation define the physical subspace
     ${\bf H}_{phys}\subset{\bf H}$.

     Note that the different classical forms same constraints
     can lead to the different operator equations.
     Indeed, as the general situation we have
     $$\Upsilon({\cal D}_1)\Upsilon({\cal D}_2)\not=
     \Upsilon({\cal D}_1{\cal D}_2)$$
     for arbitrary observables ${\cal D}_1$ and
     ${\cal D}_2$  so that the constraints $\Phi=0$ and
     $\Phi^2=0$, for example, lead to the different quantum
     theories.
     As the standard situation, we have some (singular)
     Lagrangian  ${\cal L}(q\dots\dot{q}\dots)$
     and the natural form of the constraints which are
     consequence of the definitions
     $p_i=\partial{\cal L}/\partial\dot{q}_i$ for the canonical momenta.
     In our case we construct the hamiltonian system without lagrangian
     formalism (in accordance with general Dirac's ideas); moreover the
     equality  $\Phi=0$ appears as some external condition.
     But the existence of the lagrangian scheme for any constraint
     hamiltonian dynamics is non-trivial problem. Moreover
     there is no such scheme for the arbitrary constraints
     \cite{Pavlov}.
     All these arguments  mean that we must postulate
     strongly the form of the constraint (\ref{mainco}) before
     quantization.

     Let us introduce the parameter $\lambda$ instead the
     parameter $\kappa$ such that
     $$\lambda =\frac{\alpha^{'} P}{\kappa}\qquad
     \left(P=\sqrt{P^2}\right)\,.$$
     Because  $\kappa\not=0$ in our theory such redefinition of the
     parameter will be correct. Note that
     the point  $\lambda =0$ corresponds to the massless particles.
     As the consequence we have the  following
     $\lambda$ - parametric representation\footnote{Let us remind that
     all evident expressions here are deduced from
     the general theory constructed in the work
          \cite{Tal1}.}
     of the constraint $\Phi=0$:
     \begin{eqnarray}
     \label{l1}
     &~&h_0 - \lambda^2= 0,\\[3mm]
     &~&\Phi_1(\lambda\,|\,{\cal B}_2, {\cal B}_3, P^2,S^2)
     \equiv - S^2\lambda^6 +  {1\over 4} \varrho_0^2\lambda^2 P^2-
     \nonumber\\[3mm]
     \label{l2}
     &~& -{1\over 2}\alpha^{'} P^3\varrho_0\lambda\biggl({\cal B}_3+
     {\cal B}_3^+\biggr)
    +{\alpha^{'}}^2 P^4\,\biggl({\cal B}_2^2\lambda^2+
     {\cal B}_3 {\cal B}_3^+ \biggr)=0,
     \end{eqnarray}
     where the following notations are stated:
     $${\cal B}_2 = {{\rm i}\over{2\pi}} \int_0^{2\pi}f(\xi)
     \int^\xi\overline f(\zeta)d\zeta  d\xi, \qquad
     {\cal B}_3 ={1\over{2\pi}} \int_0^{2\pi}|f(\xi)|^2\int^\xi
     f(\zeta)d\zeta d\xi.$$
     Symbol $\int^\xi f$ means such antiderivative of the function
     $f$ that has not zero mode in Fourier expansion.
     Let us note that  the Poisson brackets of l.h.s. of the
     equalities  (\ref{l1}) and  (\ref{l2})
     are vanished strongly that is why such representation
     of our constraint $\Phi=0$ will be correct.
     This form of the constraint $\Phi=0$ -- equalities
     (\ref{l1}) and  (\ref{l2}) -- is postulated before the
     quantization.

     Let us introduce the notations
     \begin{eqnarray}
     &~&{\bf h}=\gma^{-1}\Upsilon\left({1\over{2\pi}}\int\mid
     f(\xi)\mid^2d\xi\right)=\sum_n\mid n\mid{\bf b}^+_n
     {\bf b}_n\,,\nonumber\\[3mm]
     &~&{\bf w}_2 =
     {\gma}^{-1}\Upsilon\left({{\rm i}\over{2\pi}}\int_0^{2\pi}
     \overline f(\xi)\int^\xi f(\eta)d\eta d\xi\right)=
     \sum_{n\not=0}{\rm sgn}(n){\bf b}^+_n\,{\bf b}_n\,,\nonumber\\[3mm]
     &~&{\bf w}_3 =\gma^{-3/2}
     \Upsilon\left({1\over{2\pi}}\int_0^{2\pi}\mid
     f(\xi)\mid^2\int^\xi f(\eta)d\eta d\xi\right)=\nonumber\\[3mm]
     &~&={\rm i}\sum_{n,k>0}\sqrt{\frac{n}{k(n+k)}}\,\biggl[
     (n+k)\left({\bf b}_{-n-k}^+{\bf b}_{-n}{\bf b}_{-k} -
     {\bf b}_{n}^+\,{\bf b}_{k}^+\,{\bf b}_{n+k}\right)+\nonumber\\[3mm]
      &~&+ n\left({\bf b}_{n+k}^+{\bf b}_{n}\,{\bf b}_{-k}-
     {\bf b}_{-n}^+{\bf b}_{k}^+\,{\bf b}_{-n-k}\right)\Biggr]\,.\nonumber
     \end{eqnarray}
     In accordance      with      the      classical       expressions
      (\ref{l1}) and (\ref{l2}), the quantum equations for the physical
     states vectors
      $\mid\psi_{phys}\rangle\in{\bf H}$ have the form:
     \begin{eqnarray}
     \label{hl}
     \left({\gma{\bf h}}-\lambda^2\right)\mid\psi_{phys}\rangle &=& 0,\\
     \label{main1}
     \Phi_1(\lambda\,|\,{\bf w}_2, {\bf w}_3,
      P^2,S^2 )\vert\psi_{phys}\rangle &= &0.
     \end{eqnarray}
     Operator $\Phi_1$  must be constructed in accordance with the
     ordering rules formulated above.

     It can be verified directly that
     $$[{\bf h},{\bf w}_2]=[{\bf h},{\bf w}_3] =0,$$
     so that
     $$[{\bf h},\Phi_1(\lambda\,|\,{\bf w}_2, {\bf w}_3, P^2,S^2 )]=0.$$

    The structure of the space ${\bf H}$ means that any special
    solution of the system (\ref{hl}) - (\ref{main1}) has a form
    $$\mid\psi_{phys}\rangle=
      \mid\mu_\phi^2,s\rangle\mid\phi\rangle,$$
    where
      $\mid\mu_\phi^2,s\rangle\in{\bf H}_{\mu^2,s}$,
     $\mid\phi\rangle\in{\bf H}_b$ and
     $\mu_\phi^2=\mu^2\left(\mid\phi\rangle\right)$ in general.
     Such representation for the vector  $\mid\psi_{phys}\rangle$
     leads to following spectral tasks in the space
    ${\bf H}_b$~ $(\lambda^2=\gma N)$:
     \begin{eqnarray}
     \label{hll}
     &~&\left({{\bf h}}-N\right)\mid\phi\rangle=0,\\[3mm]
     \label{main2}
     &~&\Biggl[\mu^4\left(N{\bf w}_2^2+{1\over 2}({\bf w}_3^+{\bf w}_3+
     {\bf w}_3{\bf w}_3^+)\right)+\nonumber\\[3mm]
     &~&+{\epsilon\mu^3}\sqrt{\frac{N}{4\gma^3}}\biggl({\bf w}_3+
     {\bf w}_3^+\biggr)+\frac{\mu^2 N}{4\gma^3}- s(s+1)N^3
     \Biggr]\vert\phi\rangle=0,
     \end{eqnarray}
     $${\rm where}\qquad\epsilon={\rm sgn}\,\lambda =\frac{P^0}{|P^0|}.$$
     In our case $P^2>0$ that is why the value  $\epsilon$
     will be additional Cazimir function (for example we must choose
     $\epsilon =1$ for string sector $\kappa>0$).

     The eigenvalues  $N$ of the spectral task (\ref{hll})
     will be integer: $ N=0,1,2,\dots.$ Correspondent
     eigenspaces ${\bf H}_N\in{\bf H}_f$ are spanned on the
     vectors
       $$\vert l_1,\dots,  l_m\rangle=
      c_{[l]}\prod_{i=1}^m{\bf b}^+_{l_i}\mid 0\rangle\qquad
    m=1,\dots,N$$
     such that   $\sum_{i=1}^m\vert   l_i\vert=N$
    (the factor $c_{[l]}$ -- is the normalization  factor).

     The following decomposition takes place:
     $${\bf H}_b=\mathop\bigoplus_{N=0}^\infty
     {\bf H}_N,\qquad {\rm where}
     \qquad{\bf H}_0=\{c\mid 0\rangle\}.$$
     As regards the equality (\ref{main2}) it must be
     considered as the spectral task ($\mu$ will be spectral parameter)
     for each value $s=0,1/2,1,3/2,\dots$.

     The evident expressions for the operators ${\bf w}_2$ and ${\bf w}_3$
     prove that the following inclusions are true:
     $${\bf w}_2{\bf H}_N\subset{\bf H}_N,\qquad
     {\bf w}_3{\bf H}_N\subset{\bf H}_N\,.$$
     This fact means, firstly, that the expression
     $\Phi_1({\bf w}_2, {\bf w}_3,\dots)$ defines correctly
     some operator in the each space  ${\bf H}_N$ and, secondly,
     the eigenvectors $\mid\phi_\mu\rangle$ of the task
     (\ref{main2}) must be seached as
     $$\mid\phi_\mu\rangle=\mid\phi_{N,\mu}\rangle\in{\bf H}_N.$$

     It is clear that the number $d_N={\rm dim}\,{\bf H}_N$ is the
     finite number so that the operator
     $\Phi_1({\bf w}_2, {\bf w}_3,\dots)$ defines some matrix
     $d_N\times d_N$ in the each space ${\bf H}_N$.
     Thus the spectral task  (\ref{main2}) is reduced to
     the series of the matrix tasks marked by number
      $N=1,2,\dots.$

     Let $\{\mid N,l\rangle$; $l=0,1,\dots,d_N\}$ denote the set of
     the vectors  $\vert l_1,\dots,l_m\rangle$ introduced above
     and numbered in some way.
     Then the eigenvalues
      $\mu =\mu_l(s; N,\gma,\epsilon)$
     will be the roots of the algebraic equations
     ($N=1,2,\dots$):

    \begin{equation}
     \label{aleq}
     \det\Biggl[\mu^4{\bf A} +
    {\epsilon\mu^3}\sqrt{\frac{N}{4\gma^3}}{\bf B}
      +\biggl(\frac{\mu^2 N}{4\gma^3}-s(s+1)N^3\biggr)
      1_{d_N}\Biggr]=0,
     \end{equation}
     where $1_{d_N}$ is the unit matrix $d_N\times d_N$ and the
     symbols ${\bf A}$ and ${\bf B}$ denote the matrices
     \begin{eqnarray}
     {\bf A} &=& \langle l^\prime,N\mid\biggl(N{\bf w}_2^2+ {1\over 2}
      \left({\bf w}_3^+{\bf w}_3+{\bf w}_3{\bf w}_3^+\right)
      \biggr)\mid N,l\rangle,\nonumber\\[3mm]
     {\bf B} &=& \langle l^\prime,N\mid
     \biggl({\bf w}_3 + {\bf w}_3^+\biggr)\mid N,l\rangle.\nonumber
     \end{eqnarray}
     The following statement is true.

     \begin{em}
     For each numbers $N>0$, $s>0$ equation   (\ref{aleq})
     has the real positive roots, at least in some neighbourhood
     ${\cal O}$ of the point $\varrho_0=0$.
     \end{em}

     We drop out detail proof here and restrict himself by the
     concrete examples.
     Note that the degenerated case of our theory
     $\varrho_0=0$  leads to the  quasi-linear
    (in the plane  $(\mu^2, s)$) Regge trajectories
     $$s(s+1) =\alpha^2_k\mu^4, \qquad \mu^2\ge 0,
     \quad s= 0, 1/2, 1, 3/2,\dots.$$
     Corresponding asymptotics have a different slopes $\alpha_k$
     and same intersepts  $s=-1/2$.
     It is interesting that the analogous behaviour --
     the interseption of the infinitely many trajectories
     in this point -- was founded many years ago in some
     realistical model of the potential scattering
     \cite{GriPom}.

     Let the symbol $\mu{(s; N, k)}$ denotes $k$-th real root of the
     equation (\ref{aleq}) for some fixed numbers $N$ and $s$.
     It is clear that the total set of the roots will be
     enumerable thus we have the sequence $\mu_i^2(s)$
     which is need for the construction of the total space
     of quantum states ${\bf H}$.  The ''physical``  states
     $|\psi_{phys}\rangle\in {\bf H}_{phys}$
     have a form
     $$\mid\psi_{phys}\rangle= \sum_{n,s}c_{n,s}
     \mid\mu^2_{n}(s),s\rangle\mid\phi_{(n)}\rangle,$$
     where $\mid\phi_0\rangle=\mid 0\rangle,$~~
     $\mid\phi_{(n)}\rangle$ ($n=1,\dots,\infty$) --
     the eigenvector of spectral task (\ref{aleq})
     which corresponds to the eigenvalue  $\mu_{n}(s)$
     and the arbitrary constants $c_{n,s}$ must be subjected
     to the condition
     $\sum_{n,s}\mid c_{n,s}\mid^2<\infty.$

    \subsection*{IV. Examples.}
     \paragraph~

     {\bf 1.} $N=0.$  The equation (\ref{main2}) has a unique
     solution  $|\phi\rangle=|0\rangle.$ In accordance with
     definition of the number $N$ we have $\mu=0.$  Spin $s\not=0$
     in general. We can include or not include in the definition
     of the Hilbert space  ${\bf H}$ corresponding subspaces
     ${\bf H}_{0,s}$,  $s\not=0.$

     {\bf 2.} $N=1.$ The space ${\bf H}_1$ will be two-dimensional;
     it is spanned on the vectors  ${\bf b}^+_{1}|0\rangle$ and
      ${\bf b}^+_{-1}|0\rangle$. As regards to the equation
     (\ref{main2}), the operator ${\bf w}^2_2$ will be the unit
     operator (other operators will be zero). As the consequence
     we have  the equation
     $$\mu^4+\frac{\mu^2}{4\gma^3}=s(s+1).$$
     Thus we have the  non-linear Regge trajectory in the
     plane  $(\mu^2, s)$.
     Correspondent asymptotic line is following:
     $$s=\mu^2+{1\over 2}\left(\frac{1}{4\gma^3}-1\right).$$
     The results of the numerical calculations for
     $\alpha^{'}=0.9\,Gev^{-2}$ and some values of the constant
     $\gma$ are represented in the table \ref{tab:n1}.
     We can compare these masses with the masses of some
     neutral mesons with zero isospin.
     \begin{table}[h]
      \hspace{10mm}
\begin{tabular}{|c|c|c|l|}
     \hline
spin~$s$ & $\quad {M}\,(Gev)\quad$ & $\quad {M}\,(Gev)\quad$
& Some mesons\\
~~~~ & $\quad(\gma=0.44)\quad$ & $\quad(\gma=15)\quad$
& (isospin $I =0$)\\
     \hline
1 & 0.796 & 1.236 & $\omega(0.783)$; $f_1(1.285)$\\
     \hline
2 & 1.242 & 1.650 & $f_2(1.270)$; $f_2(1.640)$\\
     \hline
3 & 1.600 & 1.962 & $\omega_3(1.670)$; $\phi_3(1.850)$\\
     \hline
4 & 1.900 & 2.230 & $f_4(2.220)$\\
     \hline
\end{tabular}
     \caption{\ Mass spectrum in the sector $N=1$
         ($I=0$,  $Q=0$) }
     \label{tab:n1}
     \vspace{5mm}
     \end{table}

   {\bf 3.} $N=2.$  Corresponding subspace  ${\bf H}_2$ will be
     the linear combination of the vectors
   $${\bf b}^+_{-2}|0\rangle,\quad {\bf b}^+_{-1}{\bf b}^+_{-1}|0\rangle,
   \quad {\bf b}^+_{1}{\bf b}^+_{-1}|0\rangle,\quad
   {\bf b}^+_{1}{\bf b}^+_{1}|0\rangle,
   \quad {\bf b}^+_{2}|0\rangle,$$
   so that  ${\rm dim}\,{\bf H}_2=5.$
   The numerical results of the solution of the equation
    (\ref{aleq}) are represented in the table \ref{tab:n2a}
    (the values of the fundamental constants are
   $\alpha^{'} =0.9\,Gev^{-2}$ and $\gma = 15$):
   \begin{table}[t]
      \hspace{10mm}
\begin{tabular}{|c|c|c|c|l|}
     \hline
spin~$s$ & ${M}_1\,(Gev)$ & ${M}_2\,(Gev)$& ${M}_3\,(Gev)$
& Some mesons\\
     \hline
1 & 1.17 & 1.47 & 2.51 & $h_1(1.17)$; $f_1(1.51)$\\
     \hline
2 & 1.56 & 1.93 & 3.30 & $f_2(1.64)$; $\eta_2(1.87)$\\
     \hline
3 & 1.86 & 2.30 & 3.92 & $\phi_3(1.85)$\\
     \hline
4 & 2.11 & 2.61 & 4.49 & $f_4(2.20)$\\
     \hline
\end{tabular}
     \caption{\ Mass spectrum in the sector $N=2$  for $\gamma^{'}=15$}
     \label{tab:n2a}
     \vspace{5mm}
     \end{table}

 Thus we have three quasi-linear Regge trajectories in the Plane
  $(M^2, s)$:
   $$s \simeq \alpha_i M^2 - 0.5,\qquad i=1,2,3, $$
     where  $\alpha_1 = 1.1\alpha^{'}$,
     $\alpha_2 = 0.72\alpha^{'}$ and  $\alpha_3 = 0.24\alpha^{'}$.
 Trajectories corresponded to the index   $i=1,2$
are quite appropriate for the description of some
     standard meson states. As regards of the trajectory
     corresponded to the index $i=3$, it has the slope
     which differs essentialy from the standard one.
  In our opinion the corresponding states can be interpretted
  as some glueball states.
   Indeed, for example, the phenomenology of the scalar and tenzor
     glueballs was investigated in the work
   \cite{Burako}; it was supposed that the slope for glueball
     Regge trajectory  $\alpha_{glue}\simeq 0.2\,Gev^{-2}.$)
    \footnote{The string models of the glueballs are investigated too.
    So, in the work \cite{Solovi}
    the glueballs was considered as some states of the
     (elliptic) Nambu-Goto string.}

    It is interesting to see the mass spectrum deformation if we
     decrease the dimensionless constant $\gma$ (remember this
    procedure means increasing the  constant $\varrho_0$
     which is in-put constant for the action  (\ref{action})).
    Corresponding results for  $\gma =0.44$ are represented in the
     table \ref{tab:n2b}.
   \begin{table}[t]
      \hspace{10mm}
\begin{tabular}{|c|c|c|c|c|c|}
     \hline
spin~$s$ & $M_1\,(Gev)$ & $M_2\,(Gev)$& $M_3\,(Gev)$
& $M_4\,(Gev)$ & $M_5\,(Gev)$\\
     \hline
1 & 1.037 & 1.042 & 1.31 & 1.38 & 1.78\\
     \hline
2 & 1.441 & 1.443 & 1.81 & 1.86 & 2.67\\
     \hline
3 & 1.753 & 1.755 & 2.20 & 2.24 & 3.37\\
     \hline
4 & 2.016 & 2.017 & 2.52 & 2.56 & 3.95\\
     \hline
\end{tabular}
     \caption{\ Mass spectrum in the sector $N=2$  for $\gamma^{'}=0.44$}
     \label{tab:n2b}
     \vspace{5mm}
     \end{table}

   We can see two effects here: firstly, the trajectories
   become more non-linear for small values masses and spin
  and, secondly, some trajectories are splitting.
  Note that the existence of the mesons which have
  same quantum numbers and the near-by masses
  is considered as a problem in some works
     (see, for example,  \cite{Landsb2}).

   The operator ${\bf h}$ is the energy for the quantized
  bosonic field $b(\xi^0,\xi^1)$. Therefore, the quantum number $N$
  plays the role of internal energy level.  Note that the non-linearity
  of the Regge trajectories is increased for large $N$, just as a
  calculation difficulties,  unfortunately.

    \subsection*{V. Concluding remarks.}
     \paragraph~~~

It is our belief that further studies of suggested theory may well
help to guide the constructing of the realistic models of particles.
Indeed, the  case considered in this work connects the mass
 and the spin of extended particle with two annulators of the Poisson
 structure $\{\cdot,\cdot\}^0$  --  the  Cazimir  operators $P^2$ and
$w^2$.  In general case of our theory, when
the variables $j_a(\xi)\not=0$, the Poisson structure $\{\cdot,\cdot\}^0$
has four annulators. Additional annulators here the value
     $j_0^0=\int_0^{2\pi}j_0(\xi)d\xi$ and the annulator for the
  current algebra (\ref{jajb}) (see \cite{TakFad}). Last one
     is the value $n=(1/\pi){\arccos}[{\rm Tr}{\cal M}/2]$,
 were the matrix  ${\cal M}$ is the monodromy matrix
   of the auxiliary linear system
     $$U^{'}(\xi)+{i\over 2}\left(\sum_{a=1}^3 j_a(\xi)
     \sigma_a\right)U(\xi)=0\,.$$
Note that the standard boundary conditions on the original string
variables $X_\mu$ and $\Psi_\pm$ mean that the number $n$ will be
integer in the classical theory.
The quantization of the variables $j_a$ must be fulfilled in
     terms of the massless fermionic fields \cite{Tal4} by
     means of the bosonization prosedure \cite{Wit}.
In our opinion we can interpret the value $j_0^0$ as some
 (''generalized``)
 charge and the value $n/2$ as the isotopical spin after quantization.
The subsequent works will be devoted to the corresponded models.
The author hope, in particular, that the consideration of more
general case allows to fix the in-put constants $\alpha^{'}$
and $\varrho_0$.

 We construct the model of the free extended particle here.
The interaction of such particles is separate problem.
As regards to the possible approaches it must be remembered
the old point of view \cite{Mandel} which means that the Regge
trajectories  determine the  scattering amplitude.
Note also the work \cite{ChGoGo}, where so-called $q$-deformed dual
string theory was studied; it was shown that the $q$-deformation of
the amplitude leads to the non-linear (square-root) trajectories.

    \subsection*{VI. Acknowledgements.}
     \paragraph~~~

I am thankful to B. M. Barbashov for useful discussions on a
string theory and V. P. Pavlov for valuable advices regarding
the hamiltonian description of the model.

  \end{document}